# Cu Intercalation-stabilized 1T' MoS$_2$ with Electrical Insulating Behavior

Huiyu Nong[1], Junyang Tan[1], Yujie Sun[1], Rongjie Zhang[1], Yue Gu[1], Qiang Wei[1], Jingwei Wang[1], Yunhao Zhang[1], Qinke Wu[1], Xiaolong Zou[1], Bilu Liu[1,*]

[1]Shenzhen Geim Graphene Center, Shenzhen Key Laboratory of Advanced Layered Materials for Value-added Applications, Tsinghua-Berkeley Shenzhen Institute and Institute of Materials Research, Tsinghua Shenzhen International Graduate School, Tsinghua University, Shenzhen 518055, P.R. China

Correspondence should be addressed to B.L. (bilu.liu@sz.tsinghua.edu.cn)

## Abstract

The intercalated two-dimensional (2D) transition metal dichalcogenides (TMDCs) have attracted much attention for their designable structure and novel properties. Among this family, host materials with low symmetry such as 1T' phase TMDCs are particularly interesting because of their potentials in inducing unconventional phenomena. However, such systems typically have low quality and poor stability, hindering further study in the structure-property relationship and applications. In this work, we intercalated Cu into 1T' MoS$_2$ with high crystallinity and high thermal stability up to ~300 °C. We identified the distribution and arrangement of Cu intercalators for the first time, and the results show that Cu occupy partial of the tetrahedral interstices aligned with Mo sites. The obtained Cu-1T' MoS$_2$ exhibits an insulating hopping transport behavior with a large temperature coefficient of resistance reaching -4 ~ -2 %·$K^{-1}$. This work broadens the artificial intercalated structure library and promotes structure design and property modulation of layered materials.

# Introduction

Two-dimensional (2D) transition metal dichalcogenides (TMDCs) have attracted much attention due to their novel properties, such as topological insulator and charge-density-wave states.[1-4] The van der Waals (vdW) layered structure of 2D TMDCs enables them to accommodate various intercalators without destroying host structures.[5,6] The intercalators such as alkali ions, metal atoms, and organic molecules, can purposefully modify the material properties through charge transfer, functionalization, and interlayer decoupling.[7-9] With the combination of diverse host materials and intercalators, unexpected physical phenomena and interesting properties are observed in the intercalated 2D TMDCs. For example, Fe-intercalated $TaS_2$ shows giant magnetic coercivity in 2D thickness.[10] Imidazole-intercalated $NbS_2$ exhibits Ising superconductivity in bulk crystals.[11] Thus, the intercalated 2D TMDCs can provide a new platform for constructing artificial structures and designing properties for fundamental research and applications.

Most intercalated 2D TMDCs are based on the high symmetry octahedral (1T) or trigonal prismatic (2H) polymorphs as the hosts.[12] The intercalation of low-symmetry 1T' phase (*e.g.*, 1T' $MoS_2$), is still in its infancy despite its importance. The intercalation into 1T' TMDCs provides a compelling platform to explore due to the distinctive distorted structure and the low symmetry of the host 1T'. These intrinsic features present unique opportunities for symmetry breaking and the emergence of novel properties.[13,14] The intercalation also has the potentials to modulate the superconductivity,[15] topological nontriviality,[16] ferroelectricity,[17,18] and non-linear optical properties of 1T' TMDCs.[19] Most current studies use alkali ions, specifically $Li^+$ ions due to their smallest size, as the intercalators into the gap of 1T' TMDCs. This facilitates interlayer decoupling in bulk crystals and modulates their superconducting behaviors.[2,15,20] Recently, Zhai *et al.* realized the transition from semimetal 1T' $WS_2$ to semiconducting 1T'$_d$ $WS_2$ through proton intercalation.[21] Chen *et al.* confined zero-valent metal clusters within 1T' $MoS_2$ layers and prepared Pt, Ru, Au, and Pd clusters-intercalated $MoS_2$, improving hydrogen evolution reaction performance.[22] Thus, the exploration of intercalated 1T' TMDCs is of significance to design the properties and facilitate their applications, such as metal-semiconductor contacts[23] and catalysis.[22] However, the intercalated 1T' TMDCs suffer from poor stability[24] and low phase-purity,[22,25] hindering further study and their application potentials. Besides, the direct observation of the intercalator occupation is elusive, requiring the selection of a suitable intercalator. This is crucial for the in-depth understanding of the structure-property relationship. Interestingly, transition metal Cu, with a suitable atomic number (Z) contrast in microscopy imaging, was found to be able to tune the electrical behavior in layered materials,[9] and the coordination of Cu atom arises Jahn-Teller distortion lowering symmetry.[26,27] Therefore, Cu might be a suitable intercalator in 1T' TMDCs which can not only modify host's property, but also provide a valuable platform to identify interclation site at an atomic level.

In this work, we construct Cu-intercalated 1T' $MoS_2$ with high stability and electrical insulating behavior. The intercalation was performed using mild wet-chemistry intercalation on 1T' $MoS_2$ grown by low-pressure chemical vapor deposition (LP-CVD). We found that Cu intercalation can stabilize the 1T' structure at high temperatures up to 300 °C in air. Owing to the good stability of the intercalated materials and the suitable Z-contrast of Cu intercalators, we clearly identified the atomic structure of Cu-1T' $MoS_2$. The distribution and chemical states of Cu intercalators in 1T' $MoS_2$ were characterized by energy dispersive spectroscopy (EDS) and electron energy loss spectroscopy (EELS), respectively. Combining high-angle annular dark-field scanning transmission electron microscopy (HAADF-STEM) imaging and simulations, we found that Cu atoms tend to occupy partial of the tetrahedral interstices aligned with Mo sites to further induce disorder in the 1T' $MoS_2$. Electrical transport experiments in Cu-1T' $MoS_2$ reveal its insulating transport behavior with a large temperature coefficient of resistance (TCR) up to -4 ~ -2% $K^{-1}$, among the largest in 2D materials. This work expands the library of artificial intercalated structures and may help fabricate intercalated materials with designed functionalization and properties.

## Results and Discussion

The intercalation of Cu into 1T' MoS$_2$ is schematically shown in Figure 1a. In brief, we first grow 1T' MoS$_2$ on a c-sapphire (α-Al$_2$O$_3$) substrate by a LP-CVD method (see Methods, Figure S1). The lateral size of the 1T' MoS$_2$ belts reaches 50 μm with a thickness at the nanometer level (Figure S1c). After growth, the Cu atoms were intercalated using a mild Cu$^+$ complex solution via a wet-chemistry intercalation method under ambient condition (see Methods, Figure 1a and Figure S2). The disproportionated reaction of Cu$^+$ in the solution drives Cu$^0$ to intercalate into the vdW gaps and leaves Cu$^{2+}$ in the solution.[28] X-ray diffraction (XRD) patterns in Figure 1b show that the (003) peak of 1T' MoS$_2$ shifts from 14.82° to 14.03° after Cu intercalation, indicating that the interlayer spacing is expanded from 6.0 Å for 1T' MoS$_2$ to 6.3 Å for Cu-1T'MoS$_2$. Atomic-force microscopy (AFM) results show the morphology of 1T' MoS$_2$ does not change obviously after intercalation (Figure 1c and Figure S3). The thickness slightly increases after intercalation for both thin 1T' MoS$_2$ flake (1.57 nm, 2L, Figure 1c) and thick 1T' MoS$_2$ flake (9.7 nm, ~13L, Figure S3a-b), which matches with XRD results. High-resolution AFM results exhibit the representative Mo-Mo dimer chain, indicating the surface structure of Cu-1T' MoS$_2$ still maintains the quasi-1D chain-like 1T' arrangement after the intercalation (Figure S3).

Raman spectra exhibit characteristic change after Cu intercalation (Figure 1d). The as-grown 1T' MoS$_2$ shows Raman features at 146, 206, 276, 323, and 401 cm$^{-1}$.[24] After Cu intercalation, the out-of-plane vibration of 1T' MoS$_2$ at 401 cm$^{-1}$ is suppressed, while new Raman features emerge at 158 cm$^{-1}$, 260 cm$^{-1}$ and 315 cm$^{-1}$ in Cu-1T' MoS$_2$. Note that the peaks are different from those of other Cu-Mo-S compounds (e.g., Cu$_x$Mo$_6$S$_8$ and Cu$_2$MoS$_4$), suggesting the formation of a different? structure.[29-32] As a comparison, 1T' MoS$_2$ immersed in a pure acetone (without Cu$^+$ complex) remains unchanged in Raman spectra (Figure S4). Using Raman features as indicators, we compared the thermal stability of 1T' MoS$_2$ and Cu-1T' MoS$_2$ flakes through 30-min annealing in air at temperatures ranging from 50 °C to 300 °C. Figures 1e and 1f show the Raman spectra of 1T' MoS$_2$ and Cu-1T' MoS$_2$ flakes, respectively, with a similar thickness of ~5 nm (Figure S5). At 80 °C, 1T' MoS$_2$ starts to transform into 2H phase with Raman features of E$_{2g}$ mode at 380 cm$^{-1}$, consistent with the metastable nature of 1T' MoS$_2$.[33] In comparison, the structure of Cu-1T' MoS$_2$ remains intact exceeding 250 °C. At 330 °C, the appearance of 2H Raman features indicates its phase change. A semiquantitative description of structure retention is given by the change in Raman peak intensity $I(T)/I_0$, where $I_0$ is the intensity of the pristine sample and $I(T)$ is the intensity after annealing at different temperatures. The Raman peak at 146 cm$^{-1}$ for 1T' MoS$_2$ and the peak at 158 cm$^{-1}$ for Cu-1T' MoS$_2$ were monitored. As summarized in Figure 1g, when the temperature is above 80 °C, the Raman intensity of 1T' MoS$_2$ decreases significantly. But for Cu-1T' MoS$_2$, the Raman intensity decreases only above 300 °C. Together with its superior long-term stability (Figure S6), Cu intercalation can improve the stability of the 1T' structure of MoS$_2$.

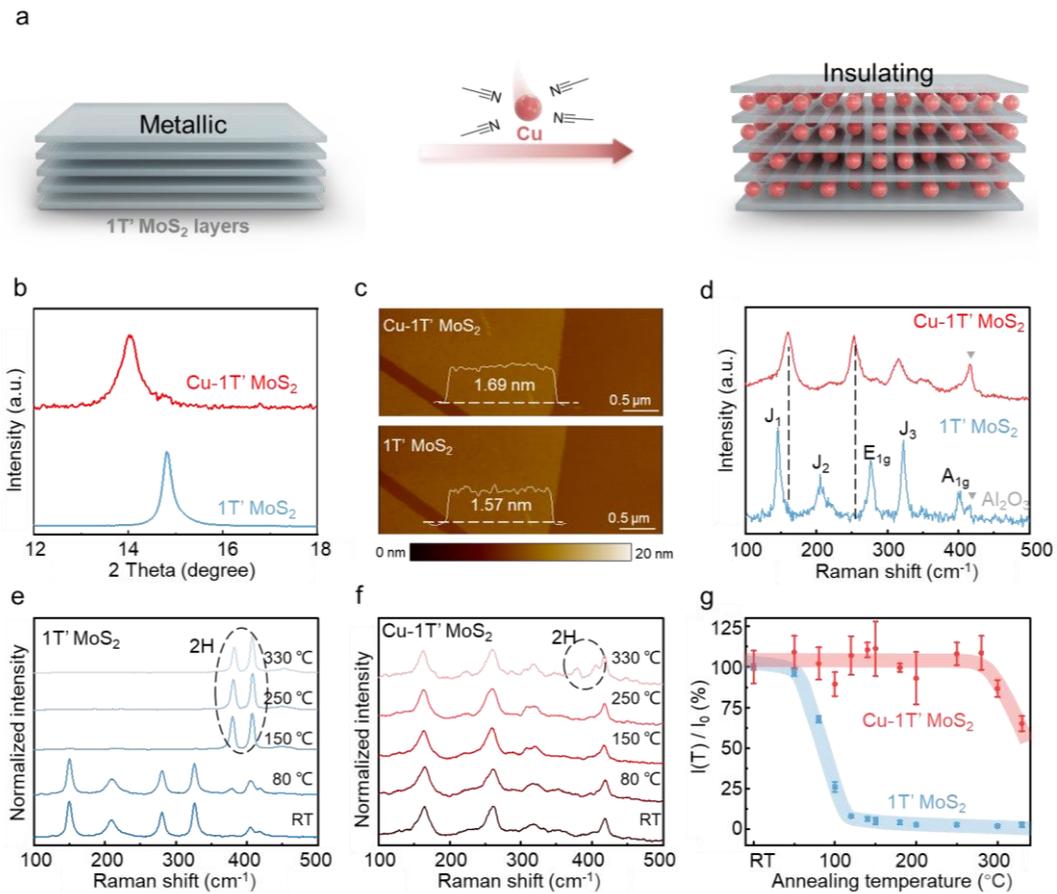

**Figure 1. Cu intercalation in 1T' MoS$_2$ and its stability.** (a) Schematic of Cu intercalation in 1T' MoS$_2$. (b) XRD of 1T' MoS$_2$ and Cu-1T' MoS$_2$. (c) AFM image of 2L 1T' MoS$_2$ w/o and w/ Cu intercalation. (d) Raman spectra of 1T' MoS$_2$ before and after Cu intercalation. (e) Raman spectra of 1T' MoS$_2$ and (f) Cu-1T' MoS$_2$ after air annealing at different temperatures for 30 min. (g) Structure retention of 1T' MoS$_2$ and Cu-1T' MoS$_2$. Here $I_0$ is the intensity of pristine sample and $I$(T) is the intensity after annealing at different temperatures. The intensity belongs to the peak at 146 cm$^{-1}$ for 1T' MoS$_2$ and the peak at 158 cm$^{-1}$ for Cu-1T' MoS$_2$.

We then study the chemical state of Cu and crystal structure of the intercalated materials. Chemical analysis was conducted to characterize the distribution and existence form of Cu intercalator using EDS and EELS (Figure 2a). EDS shows an atomic ratio of Cu of ~8% and EELS shows the characteristic step-like features of L$_3$ and L$_2$ edge of Cu, which is assigned to the zero-valent state of Cu (Cu$^0$).[28] The in-plane EDS maps of the (001) plane reveal that Cu distributes uniformly within the flake and the cross-sectional EDS maps of the (010) plane indicate that Cu penetrates through the whole flake, rather than accumulating on the surface and forming metal clusters.

We used TEM to further characterize the effect of Cu intercalation on the crystal structure. We found that the Cu-1T' MoS$_2$ shows a similar selected area electron diffraction (SAED) pattern with 1T' MoS$_2$ (Figure 2d and Figure S7). In both cases, the interplanar spacings of the (010) and (100) planes are 0.56 nm and 0.32 nm, respectively, and no obvious expansion is observed.[24] HAADF-STEM image in Figure 2e shows a similar 1T' chain to the pristine sample, indicating that the 1T' structure is preserved after intercalation, consistent with the HRAFM results. Figure 2f shows the cross-sectional HAADF-STEM image of the (010) plane. where Mo-Mo dimers are clearly observed, and the periodic sites between 1T' MoS$_2$ layers indicate the intercalation of Cu without destroying the 1T' arrangement.

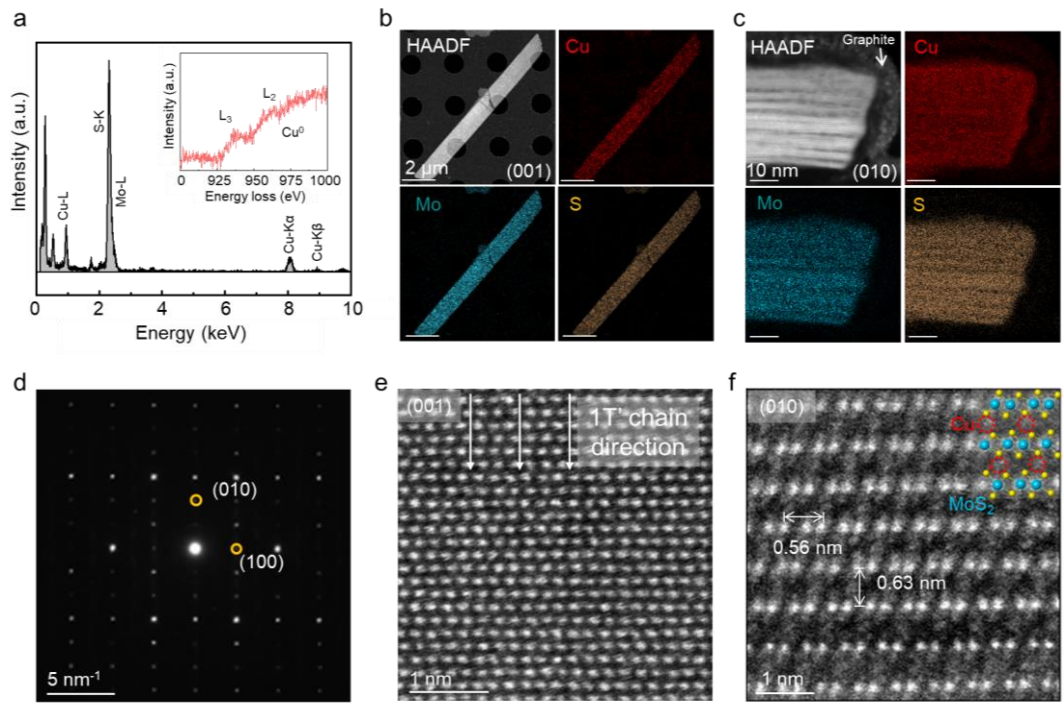

**Figure 2. Cu distribution in Cu-1T' MoS$_2$.** (a) EDS spectrum of Cu-1T' MoS$_2$, the inset is the Cu-L$_{2,3}$ EELS spectrum of Cu-1T' MoS$_2$. (b) EDS mapping of Cu-1T' MoS$_2$ along the [001] axis. (c) EDS mapping of Cu-1T' MoS$_2$ along the [010] axis. Here, a graphite flake is transferred on sample to protect the sample from possible damage during the focused ion beam cutting process. (d) SAED of Cu-1T' MoS$_2$ along the [001] axis. (e) HAADF-STEM image of Cu-1T' MoS$_2$ along the [001] axis and (f) along the [010] axis. The white arrows indicate the 1T' chain direction. The inset in (f) is the atomic model of 1T' MoS$_2$ (blue: Mo, yellow: S), showing the 1T' structure is maintained after Cu intercalation. Red-dotted circles are the Cu intercalation sites.

We further identify the Cu intercalation site at atomic level in the Cu-1T' MoS$_2$. Figure 3a shows the direct visualization of Cu lateral distribution in the (001) plane with atomic resolution. The cross-sectional HADDF-STEM image of the (010) plane of the multilayer sample suggests that Cu atoms favor aligning at Mo sites and staying in the tetrahedral interstices formed by four S atoms. Herein, we proposed two intercalation types as type I and II (Figure 3b), where the intercalation type I is half of the interstices are occupied by Cu, and type II is all the interstices are occupied by Cu. Moreover, density functional theory (DFT) calculations suggest both type I and II from the perspective of energy (see Methods, Figure S8, and Table S2).

The Z-contrast nature of the HAADF-STEM image can distinguish the pristine, type I, and type II cases through contrast differences. Here, we identified type I and type II intercalation areas in the enlarged HADDF-STEM images (Figure 3c-j), which are in good agreement with the simulation results (also see in Figure S9). It is noted that type I and type II are two succinct intercalation models identified in the extremely thin samples. As the layer number increases, the intercalation structure becomes more complicated and the real situation will be the combination of pristine, type I, and type II cases (Figure S10). The direct observation of the coexistence of type I and type II intercalation suggests that the Cu intercalators have a specific occupation of tetrahedral interstices, but they are also disordered because of the random and partial occupation of these sites.

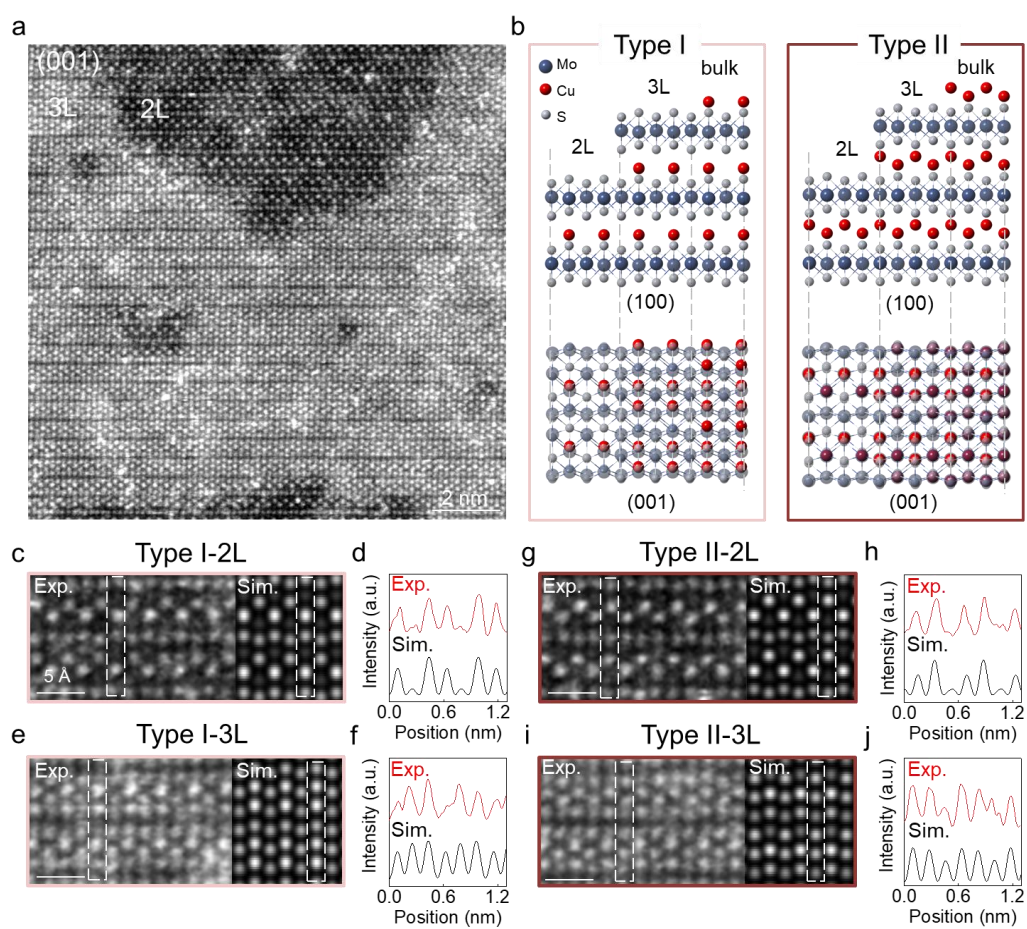

**Figure 3. Identification of Cu intercalation sites in the Cu-1T' MoS$_2$.** (a) HADDF-STEM image of the 2L/3L Cu-1T' MoS$_2$ along the [001] axis. (b) Atomic models of two intercalation structures along the [100] and [001] axes. (c) The enlarged HADDF-STEM image of the most common type I 2L area in (a) and the corresponding simulation image. Their line profiles are compared in (d). Analogous results are shown in (e, f) type I 3L, (g, h) type II 2L, and (i, j) type II 3L.

As the intercalators can strongly affect the electrical transport of host materials,[49] we fabricated devices and studied the electrical properties of Cu-1T' MoS$_2$ (Figure 4a). Four-probe electrical measurements show that the resistance of Cu-1T' MoS$_2$ is reducing with the increasing temperature (Figure 4b). This negative temperature response of resistance (d$R$/d$T$ <0) indicates the insulating behavior of Cu-1T' MoS$_2$. The temperature coefficient of resistance (TCR) is extracted from the $R$-$T$ curve by the definition formula:

$$TCR = \frac{dR}{RdT} \quad (1)$$

The TCR of Cu-1T' MoS$_2$ is -4 ~ -2 %·K$^{-1}$ at 170 ~ 300 K (Figure 4b, gray curve), which is larger than most 2D materials[34-39] and is comparable to some commercial thermistor and bolometer materials such as VO$_x$,[40-43] as shown in Figure 4c.

Meanwhile, this behavior arouses the discussion about the transport mechanism in the Cu intercalated 1T' MoS$_2$. It is well-known that 1T' MoS$_2$ is metallic at room temperature and shows superconductivity with a critical temperature T$_c$ below 10 K.[24] In other superconducting TMDCs such as NbS$_2$, Cu intercalation can improve conductivity as electron donor and tune the superconducting parameters,[44] which is different from our cases. To figure out the effect, we use the resistance curve derivative analysis to distinguish the insulating transport mechanism of Cu-1T' MoS$_2$.[45] The reduced activation energy $w$ is extracted from the $R$-$T$ curve by the defined formula:

$$w = \frac{d \ln R}{d \ln T} \quad (2)$$

The slope $p$ of ln$w$-ln$T$ curve reflects the transport mechanism (Figure 4d). For p = -1, the R-T curve can be fitted to thermal activation behavior with R∝exp(E$_a$/k$_B$T), where E$_a$ is the activation energy and k$_B$ is the Boltzmann constant. For p = -1/4, the R-T curve is fitted to three-dimensional Mott variable-range-hopping (VRH) behavior in a disordered system with $R \propto \exp(T_0/T)^{1/4}$, where $T_0$ is the critical temperature.[46] For p = 0, $w$ and $\ln w$ are constants and the R-T curve is fitted to the power-law relationship of multi-phonon hopping (MPH) behavior with $R \propto (T_0/T)^n$, where n is the number of phonons involved in hopping transport.[47] The MPH behavior is observed in a highly disordered even amorphous system with weak electron-phonon coupling.[48] Our results show that the electrical transport behavior of Cu-1T' MoS$_2$ is not band-like transport but matches well with MPH behavior (Figure 4e). The transport analysis indicates that the material system is highly-disordered and the transport is hopping through the localized states with the assistance of phonons. The number of involved phonon (n) is calculated as ~8, close to the values reported in other MPH systems such as hydrogenated amorphous $Si_{1-x}Ge_x$ with n = 5-13 and ReS$_2$-based electrical double-layer transistor with n = 5.[49,50] Similarly, Mao *et al.* observed the VRH transport in $K_x(H_2O)_yMoS_2$ (1T' MoS$_2$ intercalated with K$^+$ ions and H$_2$O molecules), suggesting that the disorder is the dominant effect of intercalation in 1T' structure.[51] In our work, the disorder effect is extended to the intercalation of zero-valent Cu atom. In addition, the coexistence of type I and II intercalation is the structural evidence of the disorder in the system. The relaxed structure in Table S2 shows that type II intercalation has ~2.8% larger lattice constant in the a-axis than type I. Thus, not only the disordered Cu atoms located between the layers but also the disordered strain within the lattice may contribute to the formation of localized states. The previous studies also demonstrated the weak localization of electrons in the disorder system is responsible for the anomalous TCR, induced by disordered microscopic strain and doping.[52-54] Overall, the Cu intercalator induces disorder in 1T' MoS$_2$, playing a major role in hopping transport and insulating behavior.

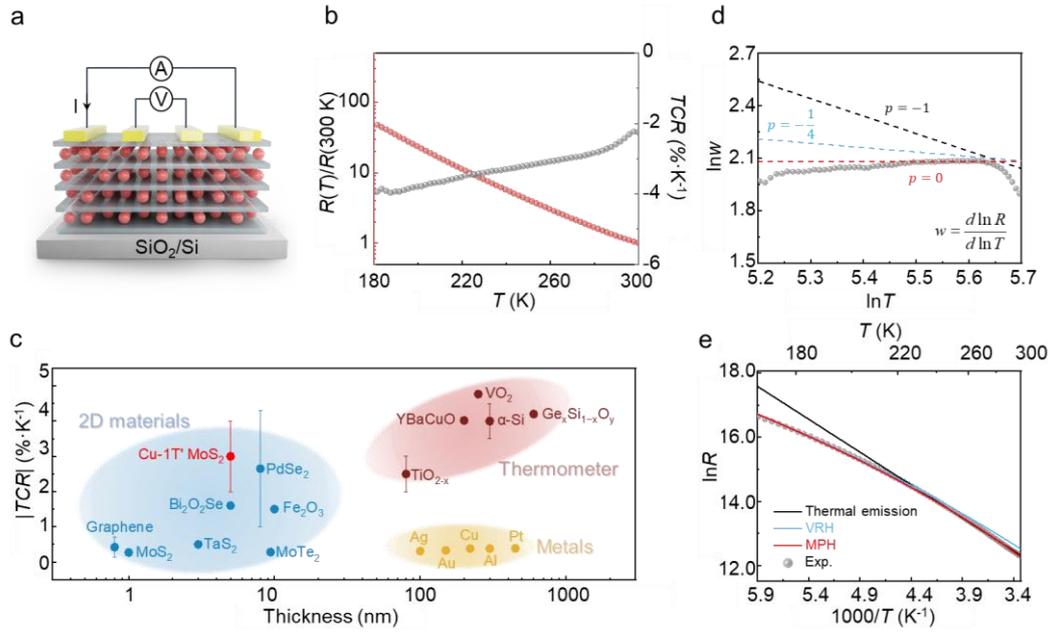

**Figure 4. Electrical transport of the Cu-1T' MoS$_2$.** (a) Schematic of four-probe electrical transport measurements. (b) Sheet resistance of Cu-1T' MoS$_2$ measured at different temperatures (red curve) by four-probe transport measurement and the derived TCR of Cu-1T' MoS$_2$ at different temperatures (gray curve). (c) The TCR comparison of Cu-1T'MoS$_2$ with other 2D materials (blue area),[34-39] commercial thermometer materials (pink area),[40-43] and metals (yellow area).[55-57] (d) The plot of ln$w$ as a function of ln$T$, $w$ is the reduced activation energy. (e) Arrenius ln$R$-1/$T$ curve fitting with different transport models. The thermal emission model can be fitted in the high temperature range of 220-300 K, the VRH model can be fitted in low temperature range of 170-220 K and the MPH model is fitted in all temperature range of 170-300 K.

## Conclusion

In this work, we constructed a stable Cu-1T' MoS$_2$ artificial structure with electrical insulating behavior. Cu$^0$ intercalates into the entire 1T' MoS$_2$, resulting in an expansion of the vdW interlayer distance. We found that Cu serves dual roles including structure stabilization and introducing disorder in 1T' MoS$_2$. Furthermore, the location and distribution of Cu intercalation in 1T' structure are clearly identified for the first time. Cu atoms occupy tetrahedral interstices, while two intercalation types of half and full intercalation coexist, suggesting the partial and random occupation by Cu atoms. Cu intercalation introduces disorder in structure, resulting in the insulating hopping transport behavior with a large TCR. This study gives insight into the intercalation manipulation of vdW materials and promotes their applications such as thermometers.

## Acknowledgements

This work was supported by the National Key R&D Program of China (2022YFA1204301), the National Science Foundation of China for Distinguished Young Scholars (52125309), the National Natural Science Foundation of China (51991343, 51991340, 51920105002, and 52188101), and the Shenzhen Basic Research Project (JCYJ20230807111619039 and JCYJ20220818101014029), Natural Science Foundation of Guangdong Province of China (2023A1515011752, 2023A1515110411), Innovation Team Project of Department of Education of Guangdong Province (2023KCXTD051), Shenzhen Science and Technology Program (ZDSYS20230626091100001), Shuimu Tsinghua Scholar Program of Tsinghua University (2022SM092), and Tsinghua Shenzhen International Graduate School-Shenzhen Pengrui Young Faculty Program of Shenzhen Pengrui Foundation (No. SZPR2023002). This work made use of the TEM facilities at the Institute of Materials Research, Tsinghua Shenzhen International Graduate School (Tsinghua SIGS).

## Additional Information

The additional supporting information could be observed from the supporting information.

## Competing Interests

The authors declare no competing interests.